\documentclass[twocolumn,showpacs,preprintnumbers,amsmath,amssymb]{revtex4}
\usepackage{graphicx,rotating}
\usepackage{epsfig}

\begin{document}
\title{Condensation Energy of the Homogeneous Electron Gas from the 
Density Functional Theory for Superconductors}

\author{M.~Wierzbowska$^{1,2}$\footnote{Present address: 
INFM DEMOCRITOS National Simulation Center, via Beirut 2--4, 34014 Trieste, 
Italy} and  J.~W.~Krogh$^{1}$}

\affiliation{ $^{1}$ Department of Theoretical Chemistry,        
University of Lund, Chemical Centre P.O.B 124, S-221 00 Lund, Sweden \\
$^{2}$ Institut f\"urn Theoretische Physik, Universit\"at
W\"urzburg, Am Hubland, D-97074 W\"{u}rzburg, Germany } 
\date{\today}

\begin{abstract}
The condensation energy of the homogeneous electron gas is calculated within
the density functional theory for superconductors. Purely electronic considerations include the
exchange energy exactly and the correlation energy on a level of the random phase
approximation. The singlet superconductivity is assumed, and the Coulomb interaction is studied 
with a model pairing potential  at the  angular momentum up to $l$=9 and at densities
1$\leq$$r_s$$\leq$10. The homogeneous gas remains  non-superconducting up to 
$r_s$$\simeq$9. Very weak negative value of the condensation energy has been found 
for f-waves and higher-$l$ pairing at $r_s$=10.
\end{abstract}

\maketitle

\section{Introduction}

For superconductivity both the electronic and the electron-phonon interactions are important.
While for the BCS \cite{BCS} superconductors the electron-phonon contributions dominate, for the novel 
high-T$_{c}$ materials the Coulomb interactions play a key role. The Eliashberg theory 
\cite{Eliashberg}, being a generalization of Migdal's theorem \cite{Migdal}, is able to treat 
the electron-phonon interaction in both the weak and the strong coupling regime, but the electronic 
interactions are averaged to a single parameter, $\mu^{\ast}$. 
This might be insufficient even for the strongly correlated systems \cite{Steward}. 
In this work, we focus on purely electronic interactions in the superconducting homogeneous gas 
for which the results obtained in last forty years still remain controversial.

The existence of superconductivity at higher angular momentum 
pairing without phononic contributions was suggested by Kohn and Luttinger \cite{Kohn-Luttinger}, 
in 1965. The mechanism proposed there was based on the presence of the long-range oscillatory 
potential in ordinary space due to the sharpness of the Fermi surface, 
and the fact that Cooper pairs \cite{Cooper}  could form taking the advantage of the attractive regions. 
Interestingly, some features of the phonon spectra have been explained due to Friedel 
oscillations \cite{Friedel}.

More than twenty years ago, Takada \cite{Takada} solved the Eliashberg equations and estimated the 
transition temperature, $T_{c}$, due to the plasmon exchange.  His solution assumed 
weak electron-phonon coupling for which the Kirzhnits, Maksimov, Khomskii approximation \cite{KMK} 
can be justified. Other authors, Rietschel and Sham \cite{RS} and Shuh and Sham \cite{Shuh}, 
solved the strong coupling Eliashberg  equations linearized in the gap function. 
For the Coulomb interactions, they also assumed the random phase approximation 
(RPA) \cite{Gell-Mann} and found unrealistically high critical temperatures. 
Later papers included vertex corrections yielding a counterpart for the plasmon exchange which
overestimate the Coulomb attraction. Obtained this way, critical temperatures were much lower 
\cite{Grabowski,Buche,T1,T2}.

Nowadays, we can use a standard solid state method, the density functional theory \cite{DFT} (DFT),
generalized to the superconducting state (SCDFT) by Oliveira, Gross and Kohn \cite{OGK} in 1988.  
The formal framework, which we briefly describe in the next section, has been developed over
more than ten years \cite{thesis} and extended for the relativistic superconductors \cite{Klaus}.
The first solutions of the SCDFT scheme for simple metals predicted the critical temperatures \cite{Fast}
quite well, although the electronic correlations have been taken into account 
only by the Thomas-Fermi screening of the Coulomb interaction. 
Then, the semi-phenomenological correlation functional was constructed and tested, 
firstly for Nb \cite{semi-Nb}, and later for YBCO \cite{semi-YBCO} using the eight-band model
\cite{8-band}.
The formulation of the local density approximation (LDA) for superconductors was given by 
Kurth {\em et al.} \cite{RPA-PRL}, in 1999. The attempt to construct the exchange-correlation 
functional from {\em first principles} was made \cite{RPA-PRL} 
using as a starting point the RPA \cite{Gell-Mann} generalized to the superconducting 
state  \cite{Rickayzen}. 
The condensation energy of the homogeneous gas has been  calculated within that scheme   
for the model pairing potential of s-type \cite{RPA-PRL} and no superconductivity 
has been  found up to $r_s$=5. 

After the discovery of the anisotropic gaps in  B2212 \cite{anisotropy-Bi} and  
YBCO \cite{anisotropy-YBCO} by the angular-resolved  photo-emission experiments, 
it is interesting to look closer at the higher angular momentum channels within the SCDFT method. 
This is a step toward DFT calculations for the superconducting state of high-T$_c$ compounds. 
Recently, the condensation energy for systems with the anisotropic gaps  has been studied 
by  Haslinger and Chubukov  within the Eliashberg theory adapted for 
the spin-fluctuation mediated pairing \cite{Chubukov-short,Chubukov-long} (see also references therein).  

The fact that, the model calculations based on results
obtained for a homogeneous gas have been performed for strongly-correlated layered superconductors 
by Kresin with co-workers \cite{applic-1} and Seibold \cite{applic-2}, shows that our studies can
contribute to understanding physics of complicated systems.
The data for a homogeneous gas obtained with our code \cite{CPC} can also give
basis for the parametrization of an LDA functional for inhomogeneous superconductors in a 
similar philosophy as an ordinary LDA is a parametrization \cite{LDA} of the quantum Monte
Carlo data \cite{QMC} calculated for a representative set of $r_s$ (Wigner radius) values. 

This paper is organized as follows: after a brief description of the theoretical background in the 
next section, we show the results of the condensation energy calculations in section III, 
and make a comparison with findings of other authors in section IV.

\section{Theoretical background}

The framework of a DFT for superconductors was formulated by Oliveira, Gross and Kohn \cite{OGK}.
It rests on 1-1 mapping between the density
\begin{equation}
n({\bf r}) = \sum_{\sigma} \langle \hat{\Psi}_{\sigma}^{\dagger}({\bf r}) 
 \hat{\Psi}_{\sigma}({\bf r}) \rangle ,
\end{equation}
and the superconducting order parameter 
\begin{equation}
\chi({\bf r},{\bf r'}) = \langle \hat{\Psi}_{\uparrow}({\bf r}) 
 \hat{\Psi}_{\downarrow}({\bf r'}) \rangle ,
\end{equation}
on one hand and the electrostatic and pairing potentials, $v_{s}({\bf r})$ and 
$\Delta_{s}({\bf r},{\bf r'})$, on the other hand.
Here, we assumed the singlet pairing, however the triplet pairing can
be treated analogously \cite{Triplet}.
The noninteracting Kohn-Sham potentials are functionals of both the normal density and the order
parameter:
\begin{eqnarray}
v_{s}[n,\chi]({\bf r}) & = & v_{0}({\bf r}) + 
\int d^{3}r' \frac{n({\bf r'})}{|{\bf r}-{\bf r'}|} + \nonumber \\
 & & v_{xc}[n,\chi]({\bf r}), \\
\Delta_{s}[n,\chi]({\bf r},{\bf r'}) & = & \Delta_{0}({\bf r},{\bf r'}) + 
\int d^{3}r' \frac{\chi({\bf r},{\bf r'})}{|{\bf r}-{\bf r'}|} + \nonumber \\
 & & \Delta_{xc}[n,\chi]({\bf r},{\bf r'}),
\end{eqnarray}
where $v_{0}({\bf r})$ is the lattice potential, and $\Delta_{0}({\bf r},{\bf r'})$ is an
external pairing potential of an adjacent superconductor. The second term in 
$\Delta_{s}({\bf r},{\bf r'})$ is the anomalous Hartree potential. The exchange-correlation
potentials are defined as functional derivatives of the xc free energy functional $F_{xc}[n,\chi]$
over the normal and the anomalous density:
\begin{eqnarray}
v_{xc}[n,\chi]({\bf r}) & = & \frac{\delta F_{xc}[n,\chi]}{\delta n({\bf r})}, \\
\Delta_{xc}[n,\chi]({\bf r},{\bf r'}) & = & -\frac{\delta F_{xc}[n,\chi]}
{\delta \chi^{\ast}({\bf r},{\bf r'})}.
\end{eqnarray}
The corresponding Kohn-Sham equations have the form of the Bogoliubov-de Gennes equations 
\cite{OGK,Suvasini} ($\mu$ is a chemical potential of the superconductor):
\begin{eqnarray}
u_{k}({\bf r}) & = & \left[ -\frac{\nabla^{2}}{2}+v_{s}({\bf r})-\mu \right] 
                     u_{k}({\bf r}) +  \nonumber \\
               &   &    \int d^{3}r' \Delta_{s}({\bf r},{\bf r'}) v_{k}({\bf r'}), \\
v_{k}({\bf r}) & = & -\left[ -\frac{\nabla^{2}}{2}+v_{s}({\bf r})-\mu \right] 
                     v_{k}({\bf r}) +  \nonumber \\
               &   & \int d^{3}r' \Delta_{s}({\bf r},{\bf r'}) u_{k}({\bf r'}),
\end{eqnarray}
and result from the diagonalization of the noninteracting Hamiltonian
\begin{eqnarray}
& & \hat{H}_{s}  =  \sum_{\sigma} \int d^{3}r \hat{\Psi}_{\sigma}^{\dagger}({\bf r})
\left[ -\frac{\nabla^{2}}{2}+v_{s}({\bf r})-\mu \right] \hat{\Psi}_{\sigma}({\bf r}) -
\nonumber \\
 & &  \left[ \int d^{3}r \int d^{3}r' \Delta_{s}^{\ast}({\bf r},{\bf r'})
\hat{\Psi}_{\uparrow}({\bf r})\hat{\Psi}_{\downarrow}({\bf r'})+H.c. \right].
\end{eqnarray}

The exchange-correlation functional includes in general the electronic and phononic contributions
\cite{thesis}. Here however, we are interested in the electronic part only, treated 
within RPA \cite{Gell-Mann} for the superconducting state \cite{Rickayzen}. 
The LDA scheme for superconductors \cite{RPA-PRL} has been constructed analogously to the local 
spin density approximation (LSDA).
In superconductors, the order parameter plays a similar role to that of the spin-magnetization in LSDA.
The electron gas is exposed to  the external pairing potential of the superconductor,
just as the LSDA gas is under the influence of a magnetic field.

\begin{figure*}
\epsfxsize=10cm
\centerline{\epsffile{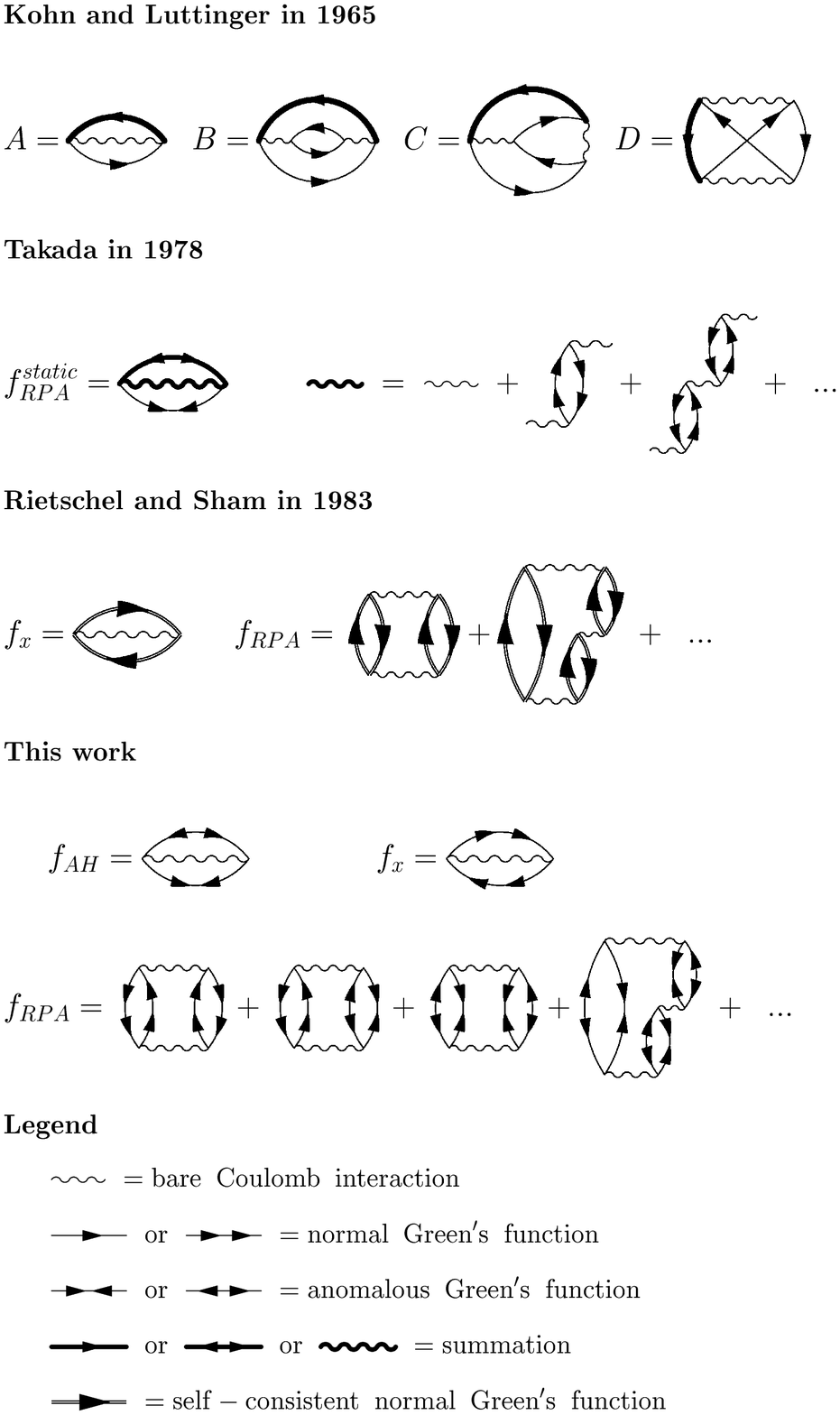}}
\caption{Feynman diagrams for the total energy contributions considered by other authors 
\cite{Kohn-Luttinger,Takada,RS} and in this work.}
\label{Feynman}
\end{figure*}

The exchange energy of the superconducting gas is given by the expression:
\begin{eqnarray}
 f_{x} [ \mu_{s},\Delta_{s}] & = &
  - \frac{1}{4}  \int \frac{d^{3}k}{(2 \pi )^{3}} \frac{d^{3}k'}{(2 \pi )^{3}}
  \frac{ 4\pi }{|{\bf k}-{\bf k'}|^{2}} \times  \nonumber \\
&&  \left[ 1- \frac {\xi_{\bf k} }{E_{\bf k}} tanh\left( \frac{\beta}{2} E_{\bf k} \right) \right]
     \times  \nonumber \\
 &&  \left[ 1- \frac {\xi_{\bf k'}}{E_{\bf k}} 
  tanh\left( \frac{\beta}{2} E_{\bf k'} \right) \right], 
\end{eqnarray}
where $E_{\bf k}=\sqrt{ \xi_{\bf k}^{2}+| \Delta_{s}({\bf k})|^{2}}$ is 
the quasi-particle spectrum and $\xi_{\bf k}=\frac{(k-k_{F})^{2}}{2}-\mu$.
The anomalous Hartree energy, $f_{AH}$, is a functional of the pairing potential only
\begin{eqnarray}
  f_{AH}  [ \mu_{s},\Delta_{s}] & = &
  \frac{1}{4}  \int \frac{d^{3}k}{(2 \pi )^{3}} \frac{d^{3}k'}{(2 \pi )^{3}}
  \frac{ 4 \pi } {|{\bf k}-{\bf k'}|^{2}} \times  \nonumber \\
&&   \frac {\Delta_{s}({\bf k})\Delta_{s}^{\ast}({\bf k'})}{E_{{\bf k}} 
  E_{{\bf k'}} }  \times  \nonumber \\
 && tanh\left( \frac{\beta}{2} E_{\bf k} \right) tanh\left( \frac{\beta}{2} E_{\bf k'} \right).  
\end{eqnarray}
The RPA energy results from the summation of bubble diagrams with normal and anomalous 
Green's functions\cite{RPA-PRL,Rickayzen}, G and F (see Fig.~\ref{Feynman}), and can be written as 
\begin{eqnarray}
   f_{RPA} [ \mu_{s},\Delta_{s}] &  = &
  \frac{1}{ 2 \beta } \int \frac{d^{3}q}{(2 \pi )^{3}} \times   \nonumber \\
&&  \sum_{\nu_{n}} \left( \log \left[ 1- \Pi_{s} ({\bf q},\nu_{n}) \frac{4 \pi}{ q^{2}} \right] \right. + \nonumber \\
&&  \left. \Pi_{s}({\bf q},\nu_{n}) \frac{4 \pi}{ q^{2}} \right),   
\end{eqnarray}
where the Matsubara frequencies, even $\nu_{n}=\frac{2n\pi}{\beta}$ and odd
$\omega_{n}=\frac{(2n+1)\pi}{\beta}$,  enter the Fourier transform of the    
irreducible polarization propagator $\Pi_{s} ({\bf q},\nu_{n})$, with ${\bf q}$ being the
momentum exchange of the interacting electrons, as follows:
\begin{eqnarray}
  \Pi_{s}({\bf q},\nu_{n}) &  = & \frac{2}{\beta} \int \frac{d^{3}k}{ (2 \pi )^{3}} 
   \times \nonumber \\
 && \sum_{\omega_{n}} [G({\bf k},\omega_{n})G({\bf k}+{\bf q},\omega_{n}+\nu_{n}) +  \nonumber \\ 
 &&   F({\bf k},\omega_{n})F^{\dagger}({\bf k}+{\bf q},\omega_{n}+\nu_{n}) ]. 
\end{eqnarray}
Evaluation of Green's functions in the polarization propagator leads to the explicit functional
of the Coulomb and pairing potentials 
\begin{eqnarray}
  \Pi_{s}({\bf q},\nu_{n})  & = & \int \frac{d^{3}k}{(2 \pi )^{3}} 
  \left\{   \frac{E_{\bf k}+E_{{\bf k}+{\bf q}}}{ \nu_{n}^{2} +
  (E_{\bf k}+E_{{\bf k}+{\bf q}})^{2}} \right. \times \nonumber \\ 
 & &  \left[ 1- \frac{\xi_{\bf k} \xi_{{\bf k}+{\bf q}}}{E_{\bf k} E_{{\bf k}+{\bf q}}} + 
  \frac {\Delta_{s}({\bf k})\Delta_{s}^{\ast}({\bf k}+{\bf q})} 
  {E_{\bf k} E_{{\bf k}+{\bf q}} } \right] \times \nonumber \\
  && \left[ \frac{1}{2} tanh\left( \frac{\beta}{2} E_{\bf k} \right) 
   + \frac{1}{2} tanh\left( \frac{\beta}{2} E_{{\bf k}+{\bf q}} \right) \right] + \nonumber \\
   && \frac{E_{\bf k}-E_{{\bf k}+{\bf q}}}{ \nu_{n}^{2} + (E_{{\bf k}}-E_{{\bf k}+{\bf q}})^{2}}
     \times \nonumber \\ 
 & &  \left[1+ \frac{\xi_{{\bf k}} \xi_{{\bf k}+{\bf q}}}{E_{{\bf k}} E_{{\bf k}+{\bf q}}} - 
   \frac {\Delta_{s}({\bf k})\Delta_{s}^{\ast}({\bf k}+{\bf q})}
  {E_{{\bf k}} E_{{\bf k}+{\bf q}} } \right] \times
   \nonumber \\
 & &   \left. \left[ \frac{1}{2} tanh\left( \frac{\beta}{2} E_{\bf k} \right)
   - \frac{1}{2} tanh\left( \frac{\beta}{2} E_{{\bf k}+{\bf q}} \right) \right] \right\} .   
   \nonumber
  \label{polar}
\end{eqnarray}
We assumed above, that the Kohn-Sham orbitals are plane waves, because we are interested in the
condensation energy of the homogeneous electron gas.
The way to obtain the LDA functional for inhomogeneous superconducting system is given in the Ref.
\cite{RPA-PRL}. Feynman diagrams for the contributions to the total energy considered in this work
are displayed in Fig.~\ref{Feynman} and compared to those diagrams from earlier papers.

The condensation energy can be obtained from 
\begin{eqnarray}
  E_{S-N} & = & T_{kin}[ \mu_{s},\Delta_{s}] - T_{kin}[ \mu_{s},0] + 
                f_{ah}  [ \mu_{s},\Delta_{s}] + \nonumber \\
          & &  f_{x} [ \mu_{s},\Delta_{s}] - f_{x} [ \mu_{s},0] + \nonumber \\
          & &  f_{c}^{RPA} [ \mu_{s},\Delta_{s}] - f_{c}^{RPA} [ \mu_{s},0],
\label{Ec}
\end{eqnarray}
where the kinetic energy difference of the superconducting and the normal state 
in a homogeneous gas at zero temperature is
\begin{eqnarray}
  && T_{kin}[ \mu_{s},\Delta_{s}] - T_{kin}[ \mu_{s},0]    =   \nonumber  \\ 
  && \sum_{{\bf k}} \frac{k^{2}}{2}   \left[ \frac{1}{2} - 
       \frac{1}{2} \frac{\xi_{\bf k}}{E_{\bf k}} 
      \right] \; sgn({\bf k} - {\bf k}_F) . \nonumber   
\end{eqnarray}

\section{The condensation energy calculations}

The first calculations of the condensation energy within the LDA for superconductors with the 
RPA functional were performed for the s-wave pairing only and no superconductivity was 
found \cite{RPA-PRL}. 
We calculate the condensation energy of the homogeneous electron gas 
at zero temperature, assuming the model non-spherical gap function of the form 
\begin{equation}
  \Delta_{s}^{lm}( {\bf k})  =  \delta \exp \left( \frac{-( k-k_{F})^{2}}
  { \sigma ^{2} } \right) P_{l}^{m}( {\bf k}),
\end{equation}
where $\delta$ and $\sigma$ are parameters in units of $\mu$ and $k_{F}$ respectively, and
$P_{l}^{m}({\bf k})$ are associated Legendre polynomials.
Above parametrization of the gap makes it possible to control the strength, 
the range and the angular shape of the pairing. 
In this work, we are mainly interested in the angular part. It will be clear from the further
discussion that, the variational determination of $\delta$ and $\sigma$ in such a way that 
the condensation energy is maximally negative would lead to either zero values for these parameters 
if $E_{S-N}$ is positive,  or to infinite values of these parameters if $E_{S-N}$ is negative. 
As we will see later, the condensation energy is also monotonic with $l$, but we were not able to predict 
this result from the analytical form of the expression (\ref{Ec}).

The s-wave calculations appeared smooth in $\delta$ and $\mu$ parameters \cite{RPA-PRL} in the range
0.01$<$$\sigma$$<$1 and 0.01$<$$\delta$$\cdot$100$<$1 at $r_{s}$=0.1 and 1$\leq$$r_{s}$$\leq$5.
Therefore in this work, we fixed the strength of the potential at $\delta$=0.01$\cdot \mu$ and
the range of the pairing interaction at $\sigma$=0.1$\cdot k_{F}$,
and we present results for this choice of the parameters.
Later, we will discuss changes in the condensation energy when it is calculated with two other sets of
parameters $\delta$ and $\sigma$, namely with $\delta$=0.01$\cdot \mu$ and $\sigma$=0.05$\cdot k_{F}$,
and with $\delta$=0.001$\cdot \mu$ and $\sigma$=0.1$\cdot k_{F}$.

Turning to the details of the implementation, the eight-dimensional integrals of the energy functionals 
have been reduced by one dimension
in electronic Matsubara frequency which can be evaluated analytically.
Several singularities present in the formulas need special grids. For the calculation of $f_{AH}$
and $f_{x}$ and the radial part of $f_{RPA}$, we used a modified Gauss-Legendre quadrature. 
For the angular part of $f_{RPA}$, we combined
the Lobatto grid \cite{math} for the {\bf q}-momentum integration, and the
Sobolev's quasi random method \cite{NR} to  generate the mesh used by the 
Monte-Carlo quadrature over the angular part of {\bf k}-momentum. 
Details of the singularities and the parallel code are given 
in Ref. \cite{CPC}.

We will focus now on the results obtained with $\delta$=0.01$\cdot \mu$ and $\sigma$=0.1$\cdot k_{F}$. 
We see in Fig.~2 the angular momentum 
dependence of the condensation energy and its components: the anomalous Hartree energy ($f_{AH}$),
and the difference (S-N) of the exchange energy $f_{x}$ between the superconducting and 
the normal state and that difference (S-N) of the RPA correlation energy $f_{RPA}$, 
calculated at $r_{s}$=1 and at $r_{s}$=10. The condensation energy and all its components decrease 
monotonically with the angular momentum. 
The anomalous Hartree energy is the biggest positive component, almost completely balanced by 
the negative RPA energy which acts in favor of superconductivity. The exchange
energy difference between the superconducting and the normal state is positive, and tends to 
destroy pairing.
We show results for all momentum numbers $l$ from 0 to 9. However for the antisymmetric fermionic 
function only the even numbers make sense, because  we assumed the singlet spin pairing for the 
order parameter. 
The condensation energies at $r_{s}$=1 are positive for 0$\leq$$l$$\leq$9. At the density 
$r_{s}$=10, the s-wave pairing also do not allow superconductivity. 

\begin{figure}
\epsfxsize=7.5cm
\centerline{\epsffile{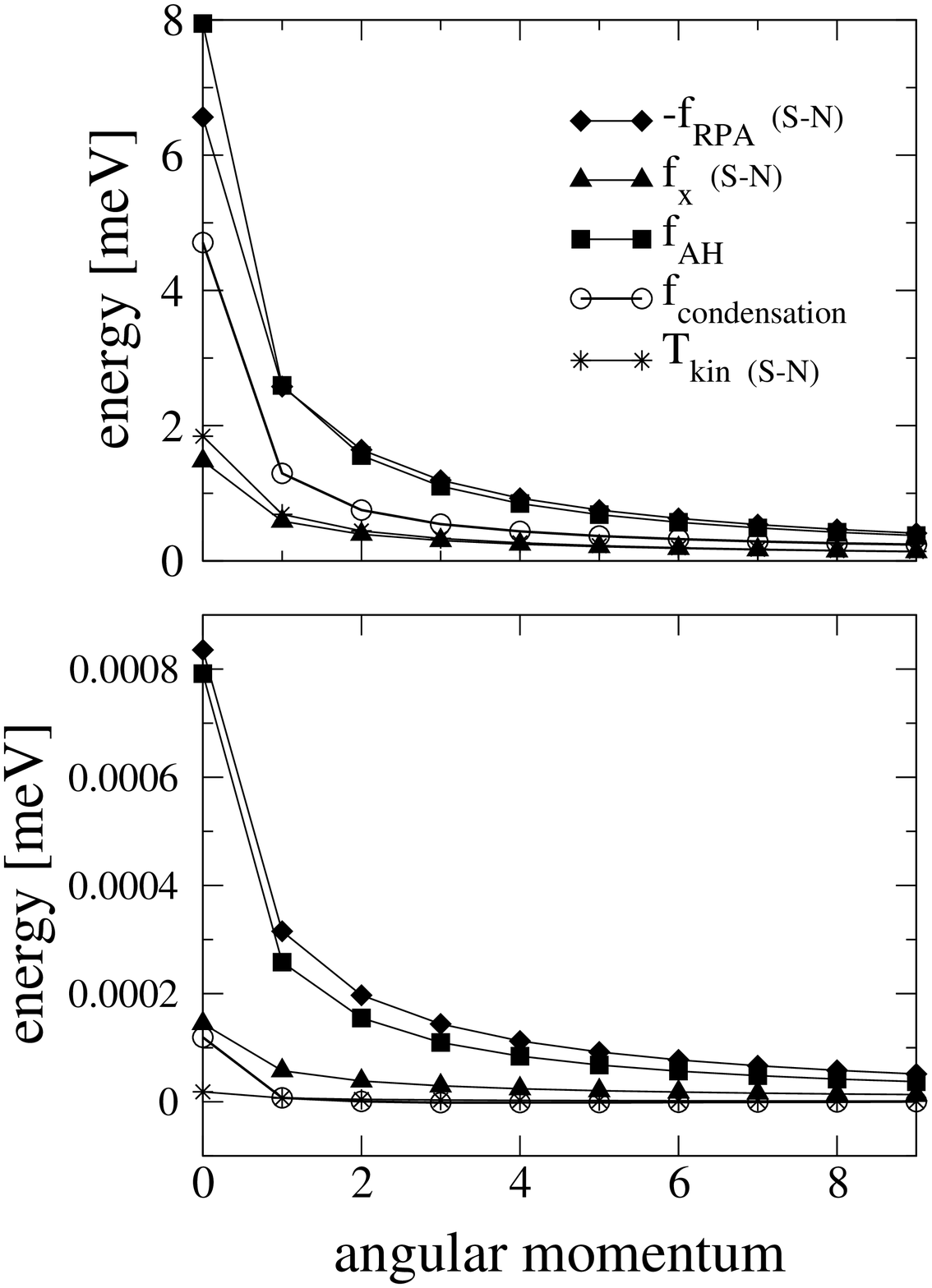}}
\caption{The condensation energy, $f_{condensation}$, and its components: the difference between 
the superconducting and the normal state of the RPA correlation energy $f_{RPA}$ (S-N), and that of  
the exchange energy $f_{x}$ (S-N), and of the kinetic energy $T_{kin}$ (S-N) and 
the anomalous Hartree energy $f_{AH}$ (this is nonzero only in the superconducting state). 
All energies are shown calculated at two densities: $r_{s}$=1 a.u. (upper panel) and 
and $r_{s}$=10 a.u. (lower panel). The model parameters in the pairing potential were fixed at 
$\delta$ = 0.01$\cdot \mu$ and $\sigma$ = 0.1$\cdot k_F$. Zero temperature was assumed.} 
\end{figure}

\begin{table}
\begin{tabular}{ccccccc}
\hline
\hline \\[-0.2cm]
  $r_{s}$ & & $f_{cond.}$ & $-f_{RPA}$  & $f_{x}$ & $f_{AH}$ & $T_{kin}$  \\[0.1cm]
\hline \\[-0.3cm]
 \multicolumn{6}{c}{s-waves} \\
\hline \\[-0.2cm]
  1   & & 4.71e-0 & 6.56e-0 & 1.48e-0 & 7.95e-0 & 1.84e-0 \\ 
  3   & & 2.30e-2 & 9.37e-2 & 1.81e-2 & 9.79e-2 & 7.59e-2 \\
  5   & & 2.82e-3 & 1.28e-2 & 2.33e-3 & 1.27e-2 & 5.90e-4 \\ 
  7   & & 6.81e-4 & 3.33e-3 & 6.05e-4 & 3.29e-3 & 1.10e-4 \\
 8.5  & & 2.87e-4 & 1.55e-3 & 2.78e-4 & 1.52e-3 & 4.16e-5 \\
 10   & & 1.19e-4 & 8.35e-4 & 1.45e-4 & 7.91e-4 & 1.84e-5 \\[0.1cm]
\hline \\[-0.3cm]
 \multicolumn{6}{c}{d-waves} \\
\hline \\[-0.2cm]
  1   & & 8.28e-1 & 1.64e-0 & 3.93e-1 & 1.56e-0 & 4.43e-1 \\ 
  3   & & 2.04e-3 & 2.18e-2 & 4.80e-3 & 1.92e-2 & 1.82e-3 \\
  5   & & 4.21e-4 & 2.82e-3 & 6.19e-4 & 2.48e-3 & 1.42e-4 \\
  7   & & 8.40e-5 & 7.49e-4 & 1.61e-4 & 6.46e-4 & 2.64e-5 \\
 8.5  & & 2.16e-5 & 3.59e-4 & 7.38e-5 & 2.97e-4 & 9.99e-6 \\
 10   & & 7.80e-7 & 1.97e-4 & 3.85e-5 & 1.55e-4 & 4.43e-6 \\[0.1cm]
\hline \\[-0.3cm]
 \multicolumn{6}{c}{f-waves} \\
\hline \\[-0.2cm]
  1   & & 7.24e-1 & 1.10e-0 & 3.02e-1 & 1.19e-0 & 3.32e-1 \\
  3   & & 3.07e-3 & 1.56e-2 & 3.69e-3 & 1.36e-2 & 1.37e-3 \\
  5   & & 3.31e-4 & 2.01e-3 & 4.75e-4 & 1.76e-3 & 1.06e-4 \\
  7   & & 5.87e-5 & 5.41e-4 & 1.23e-4 & 4.57e-4 & 1.97e-5 \\
 8.5  & & 1.21e-5 & 2.62e-4 & 5.66e-5 & 2.10e-4 & 7.48e-6 \\
 10   & & -1.45e-6 & 1.44e-4 & 2.95e-5 & 1.10e-4 & 3.32e-6 \\[0.1cm]
\hline
\hline
\end{tabular}
\caption{The condensation energy, $f_{cond.}$, and its components: 
$f_{RPA}$ (S-N), $f_{x}$ (S-N), $f_{AH}$ and $T_{kin}$ (S-N); calculated for s-waves, d-waves, 
and f-waves for fixed  $\delta$=0.01$\cdot \mu$ and $\sigma$=0.1$\cdot k_F$, and at zero temperature 
(T=0~K) for different electronic densities, $r_{s}$. Energies are given in meV, and parameters: 
$\mu$, $k_F$ and $r_{s}$ are in atomic units.}
\end{table}

\begin{table}
\begin{tabular}{ccccccc}
\hline
\hline \\[-0.2cm]
  $l$ & & $f_{cond.}$ & $-f_{RPA}$  & $f_{x}$ & $f_{AH}$ & $T_{kin}$  \\[0.1cm]
\hline \\[-0.3cm]
 \multicolumn{6}{c}{$r_{s}$=1, $\delta$=0.01$\cdot k_F$, $\sigma$=0.05$\cdot\mu$} \\
\hline \\[-0.3cm]
0   & & 3.22e-0 & 4.91e-0 & 1.09e-0 & 5.65e-0 & 1.39e-0 \\
1   & & 9.18e-1 & 2.03e-0 & 4.47e-1 & 1.97e-0 & 5.37e-1  \\
2   & & 5.39e-1 & 1.34e-0 & 3.05e-1 & 1.23e-0 & 3.52e-1 \\
3   & & 3.94e-1 & 1.00e-0 & 2.37e-1 & 8.90e-1 & 2.66e-1  \\[0.1cm]
\hline \\[-0.2cm]
 \multicolumn{6}{c}{$r_{s}$=1, $\delta$=0.001$\cdot k_F$, $\sigma$=0.1$\cdot\mu$} \\
\hline \\[-0.3cm]
0   & & 1.07e-1  & 2.24e-1 & 3.90e-2 & 2.58e-1 & 3.39e-2 \\
1   & & 2.59e-2  & 8.12e-2 & 1.44e-2 & 8.06e-2 & 1.20e-2 \\
2   & & 1.43e-2  & 4.97e-2 & 9.23e-3 & 4.72e-2 & 7.53e-3 \\
3   & & 1.02e-2  & 3.53e-2 & 6.90e-3 & 3.31e-2 & 5.53e-3 \\[0.1cm]
\hline \\[-0.2cm]
 \multicolumn{6}{c}{$r_{s}$=10, $\delta$=0.1$\cdot k_F$, $\sigma$=0.5$\cdot\mu$} \\
\hline \\[-0.3cm]
0   & & 7.85e-3  & 2.90e-2 & 5.05e-3 & 3.04e-2 & 1.40e-3 \\
1   & & 1.49e-3  & 1.03e-2 & 2.23e-3 & 9.02e-3 & 5.39e-4 \\
2   & & 1.13e-3  & 5.91e-3 & 1.58e-3 & 5.11e-3 & 3.53e-4 \\
3   & & 9.17e-4  & 4.05e-3 & 1.25e-3 & 3.45e-3 & 2.67e-4 \\[0.1cm]
\hline
\hline
\end{tabular}
\caption{The condensation energy and its components: $f_{RPA}$ (S-N), $f_{x}$ (S-N), $f_{AH}$ and $T_{kin}$ (S-N),
calculated as functions of the angular momentum $l$ for three sets of parameters: $r_{s}$, $\delta$ and $\sigma$. 
Energies are given in meV.}
\end{table}
%
%
%
%
%
%

In order to show the energetics as a function of the density of a homogeneous electron gas, 
we present, in Table~1, the condensation energy and its components for the  
s-type and the d-type as well as the f-type pairing potential, 
and for the $r_{s}$ parameter in the range of 1--10 a.u. 
Decreasing the density from  $r_{s}$=1  to  $r_{s}$=3, all energies decrease 
by two orders  of magnitude, from a few meV to a few per cent of meV. 
Further dilution of the electron gas to  $r_{s}$=10 lowers the energies by 
another two orders of magnitude. This shows how
delicate the balance is between the superconducting phase and the normal state.

Slightly negative values of the condensation energies at $r_{s}$=10 are obtained for
f-waves and higher angular momentum pairing, but all components are very small and
the most negative value is of order $\sim$1.5$\cdot$10$^{-6}$ meV.
Since the biggest contributions to this negative energy are $f_{RPA}$ and $f_{AH}$, which are of order
1$\cdot$10$^{-4}$ meV while the condensation energy is of order 1$\cdot$10$^{-6}$ meV,
and since we trust to our numerical results up to the three leading digits, it is plausible
that a change by one in the last position in the correlation and the anomalous Hartree energy may cause the
change of a sign in the condensation energy. Actually for the same reason, the condensation energy at
$r_{s}$=10 for the d-wave pairing could be negative (because it is three orders of magnitude smaller than
the leading contributions). But this uncertainty due to the numerical accuracy will happen neither for s-wave
pairing nor for smaller $r_{s}$ parameters, as one can see in Table~1.
Later, we will show calculations for $r_s$=10 with another choice of $\delta$ and $\sigma$.

Finally, we change the parameters in the model pairing potential. 
These results are presented in Table~2 for $r_s$=1 and for the angular momentum up to 3 
(since before we found the possibility for a phase transition for f-waves).

Firstly, we change $\sigma$ parameter for 0.05$\cdot k_{F}$ (before it has been fixed at 0.1$\cdot k_{F}$),
and we keep the same $\delta$ as in Table~1, i.e. 0.01$\cdot\mu$.  For this choice of $\delta$ and $\mu$, 
all energies are smaller, and we do not find superconductivity at $r_{s}$=1 up to $l$=3.
The typical band widths at the Fermi surface are much more narrow than 0.1$\cdot k_{F}$. The results of calculations
for  s-waves performed by previous authors \cite{RPA-PRL} were smooth in $\sigma$, and our results
with two values of $\sigma$ appear smooth with respect to a variation of the angular momentum.
Therefore, we do not expect any change in conclusions by changing the $\sigma$ parameter.    

Secondly, we change $\delta$ parameter for 0.001$\cdot\mu$ (before it has been fixed at 0.01$\cdot\mu$)
and we keep the same $\sigma$ as in Table~1, i.e. 0.1$\cdot k_F$. Now, all energies are smaller, 
and the decrease of the condensation energy is about one and half order of magnitude while the change of 
the pairing amplitude $\delta$ is by a factor of 0.1. 
The BCS behavior of the condensation energy is proportional to a square of the gap, $\sim \Delta^{2}$.
We find in our calculations that, the dependence of the condensation energy on the gap
amplitude is a bit weaker than the BCS one.
For the lower momentum channels a power of that dependence is smaller than for the higher angular momentum
channels. This seem to be in contrast with measurements for ordinary, BCS like, superconductors which have 
a gap of the s-type.
On the other hand, measurements are not able to split the purely electronic and the phononic contributions, and
we did not add phonons to our calculations.

Our result indicates that, we should not expect superconductivity at small $r_{s}$. Since all the energy
components grow with $\delta$ and $\sigma$, we can say that the maximally negative condensation energy would be
for $\sigma$=$k_{F}$ (all states contribute to pairing) and that there is no upper limit for $\delta$.
This is due to the fact that our condensation energy is proportional to $\Delta^{n}$ with 1.5$<$$n$$<$2.
The value of $\delta$, which we set to 0.01$\cdot\mu$, is much bigger than typical gaps.
For instance for Nb, we have s-wave pairing, $r_{s}$=0.87, $\mu$=33.13~eV, while the experimental gap is 1.55~meV. 
We have chosen large delta, i.e. 0.01$\cdot\mu$$\sim$330~meV, for most of our calculations 
for the sake of accuracy, since the conclusions about the angular dependence of the condensation energy 
do not change with this parameter. 

For lower density, i.e. $r_s$=10, we calculated again the condensation energy but for a new choice of the gap parameters:
$\delta$=0.1$\cdot$$\mu$ and $\sigma$=0.5$\cdot$$k_F$, which are unrealistically high but the numerical accuracy is
much better in this case. The result of above calculations is negative for superconductivity, 
which makes also a situation, that the phase transition occurs for a small gap, less probabilistic.

If we wanted to include the electron-phonon or electron-paramagnon interaction, the expression for the total energy
would be frequency independent because the frequency is integrated out in the SCDFT scheme (see for instance
Ref.~\cite{OGK,thesis}). Similar to the total energy, also the gap function would be static, as it is now.
This is in contrast to the Eliashberg formalism where inclusion of strong coupling changes the gap function to
a dynamic parameter \cite{Chubukov-long}.
As for the feedback effect of phonons or spin-fluctuations for the electronic energy and {\em vice versa}, 
this effect would exist if one solved the SCDFT equations in the Bogoliubov-de Gennes form 
(for this formulation of SCDFT see for example Ref.~\cite{semi-YBCO}).  
In the way how we calculate the condensation energy in this work, the aforementioned feedback
would not exist since all energy components contribute to the total energy independently through a single equation.
This is another difference from the Eliashberg scheme, where three coupled equations have to be solved for the 
self-energy $\Sigma$, pairing-vertex $\Phi$ and polarization $\Pi$ \cite{Chubukov-long,Chubukov-short}.  
  
Summarizing results: the condensation energy is positive for all angular momentums, which indicates that we are
unlikely to obtain superconductivity from the electronic correlations only.
For the very dilute gas at $r_{s}$=10, we obtained slightly negative
values of the condensation energy when the gap is small. 
These results should be however viewed with care, because all
the energy components are very small and at the limit of numerical accuracy for the multidimensional 
quadrature. We trust to three leading digits in Tables~1 and 2, while the condensation energy at $r_{s}$=10
is 3 orders of magnitude smaller than the biggest contributions.
Also the random phase approximation \cite{Gell-Mann} for the correlation energy is exact 
only in the limit of high density,\cite{Nozieres} i.e. $r_{s}$$<$1.  
This treatment of the Coulomb interaction might be insufficient for  the density at $r_{s}$=10. 
Whether it is an appropriate approach it depends also on the angular momentum of 
the pairing potential and on the physical property one is interested \cite{T2,Tspd}. 

\section{Discussion and Conclusions}

Whether the superconductivity can exist without phonons or not is a very old problem.
In 1965, Kohn and Luttinger \cite{Kohn-Luttinger} suggested a mechanism of the Cooper pair
\cite{Cooper} formation in the homogeneous gas due to Friedel oscillations \cite{Friedel}. 
These authors did not assume any particular form of the interaction, which could even be purely repulsive, 
since the attractive regions could form in real space because of a sharpness of the Fermi surface
in the reciprocal space. 
It has been discussed that, for the pairing potential at higher angular momentum, 
the superconducting state was more favorable than for s-waves.
The above conclusions were based on the mathematical analysis of irreducible vertexes with the
particle-particle interaction up to the second order (see Fig.~\ref{Feynman}). The criterion used
for the superconductivity was the occurrence of a pole at T$_{c}$ in the scattering amplitude 
for pairs of quasi-particles of equal and opposite momenta and in the total energy corresponding 
to two particles on the Fermi surface.

Later work on the superconducting homogeneous gas within the Eliashberg theory, treating the Coulomb
interactions on the RPA level and beyond, can be compared to the results presented in this paper.

Superconductivity obtained due to the plasmon exchange alone seemed to be overestimated.
Several papers solving the Eliashberg equations with the RPA model for the Coulomb interaction predicted 
superconductivity for s-wave pairing at quite high densities. Within the weak coupling limit
of the electron-phonon interaction the Eliashberg equations could be considered as 
{\bf k}-dependent only due to the KMK approximation \cite{KMK}. 
In the limit of strong coupling to phonons, the {\bf k}- and  $\omega$-dependent equations have to be 
solved. While first approximation led to superconductivity \cite{Takada} at the density lower 
than $r_s$=6, solving the strong coupling regime equations yields a change in a sign of 
the Coulomb parameter  \cite{RS} ($\mu^{\ast}$$<$0) at the density $r_s$$\sim$2.5. 
The total energy diagrams included in both approaches are shown in Fig.~\ref{Feynman}.
The common assumption in the aforementioned two approaches is that the polarization function, 
which enters the RPA screened interaction contains only normal Green's function loops, 
neglecting the anomalous ones which are also included in our scheme. 
In addition, the normal Green's function used by Sham and co-workers was obtained
self-consistently in contrast to all the other papers discussed here. The correctness of such an approach
was discussed by several authors \cite{ShamAlone,T2} also by occasion of the GW 
approximation \cite{Holm,Godby}.
The criterion for superconductivity used by Takada~\cite{Takada} was a nonzero critical 
temperature. Sham and co-workers\cite{ShamAlone,RS,Grabowski,Buche} considered the electron gas to be 
superconducting when the Coulomb pseudo-potential $\mu^{\ast}$ was positive. 

A number of papers that included vertex corrections on different level approximations predicted 
decreased temperatures of phase transition.
Grabowski and Sham \cite{Grabowski} studied the $\omega$-dependent only Eliashberg equations 
with vertex corrections up to the second order. 
Within that simplified model, the signum of $\mu^{\ast}$ changed at $r_s$=7.
More extensive study of vertex corrections were done within the KMK, {\bf k}-dependent scheme 
extended to the strongly  correlated systems by Takada \cite{T1}, who included more than 50 diagrams 
systematically using the effective-potential expansion. As a conclusion of that work
the phase transition occurred at $r_s$$>$3.9, the maximum of the critical temperature was obtained
at $r_s$=7.2, and $T_c$ decreased for lower densities.
B\"uche and Rietschel \cite{Buche} added the vertex corrections, within the phenomenological 
model by Kukkonen and Overhauser (KO) \cite{semi-vertex}, to the earlier work by 
Rietschel and Sham \cite{RS}, and did not obtain superconductivity up to $r_s$=5. 
The parameter $\mu^{\ast}$ for gas in the range of 1$\leq$$r_s$$\leq$5 was positive and varied 
between 0.05 and 0.1. Then, Takada \cite{T2} performed {\bf k}- and $\omega$-dependent calculations 
with the local-field correction of the KO model and showed a significant effect of corrections beyond RPA 
around $r_s$=5.  Another observation made in his work was that, although the compressibility 
$\kappa$ and the spin susceptibility $\chi$ were strongly dependent on the vertex corrections, 
$T_{c}$ of gas at the density $r_s$$>$20 was similar to the temperature obtained from RPA. 
For $r_s$$>$40 the critical temperature has been approximated by $T_c$$\approx$0.04$\cdot$E$_{F}$.

All the aforementioned papers on the superconducting homogeneous gas, within RPA and beyond, dealt with
the pairing potential of the s-type. The s-, p- and d-type pairing potentials examined 
in Ref.~\cite{Tspd} led to the conclusion that the vertex corrections to s-waves are 
much more important than to the higher momentum channels.

In this work, we calculated the condensation energy at zero temperature, instead of solving the gap
equation for finite temperatures. This way we are not able to determine the critical temperature
for the densities at which we find superconductivity. On the other hand, negative values of the
condensation energy are very small, of order $\sim$1.5$\cdot$10$^{-6}$ meV, at the densities 
$r_s$$\geq$10 for f-waves and the higher-$l$ channels. For d-waves, the condensation energy is 
positive but very small.

We concluded that this effect might be due to an accuracy of the multidimensional numerical
quadrature of objects with many singularities. On this point we want to comment that most
of standard quantum chemistry programs which calculate two-electron integrals do not accede an
accuracy higher than 6 important digits. Our task is even more difficult because in addition to 
calculating the $\bf k$- and $\bf q$-momentum vectors, we have to perform the quadrature over 
the bosonic Matsubara frequency, and singularities in the superconducting state have much more 
complicated shape \cite{CPC} than those of the two-electron integrals calculated for the normal state.

The positive aspect of the method employed here is the absence of any approximation except RPA. 
The DFT is exact for the ground state studied here. We do not make any assumption about phononic
interactions, which we neglect. But if we wanted to include the electron-phonon interactions, then
the way of treating the weak and the strong coupling would be the same \cite{Fast}. We do not drop
either the momentum dependence or the frequency dependence of the Coulomb interaction.
The vertex corrections, especially important for the moderate densities for the properties
like the critical temperature \cite{T2}, are not taken into account in this work.
However, we believe that the inclusion of vertex corrections would not change the conclusions,
because for f-waves the strength of the Coulomb attraction is not as much overestimated by the
polaron exchange as for s-waves \cite{Tspd}.

Another interesting question would be whether it is possible that at some densities s-waves are 
favorable for the superconducting state and at other densities the higher-$l$ pairing would lead 
to lower energy. Such s-wave to p-wave transition has been reported by Takada \cite{Tspd} at 
$r_{s}$=4.7 while for higher densities the energy of p-waves was lower.
K\"uchenhoff and W\"olfle \cite{p2s-trans}, by solving two
coupled Bethe-Salpeter equations for the two-particle vertex functions in the particle-hole
channels, found the p-wave superconductivity for 10$<$$r_s$$<$35 and the s-wave superconductivity
for $r_s$$>$35.
From our results, which are monotonic with the angular momentum number, such $l$-wave to $l'$-wave 
transition seems not to be the case~\footnote{Although we did not calculate the total energies 
for the normal and the superconducting state separately, but only the differences of those, 
it is still plausible to compare the condensation energies with pairing of different momentum, 
because the normal state energy is the same for all numbers $l$ (since $\Delta_{s}$=0).}.

We assumed singlet pairing. Thus, only the even numbers $l$ 
(like s-waves, d-waves etc.) are relevant, and it would be incorrect from the symmetry point of 
view to compare the total  energy of s-waves with the total energy of p-waves.
On the other hand, if there is not much energy gain by the Cooper pair formation then the
spin pairing in the superconducting phase probably is the same as the magnetic phase of the
normal state. Within the QMC methods \cite{Ortiz}, it has been widely examined theoretically 
that the ground state of the homogeneous electron gas is paramagnetic for high and intermediate
densities, and the transition to the ferromagnetic phase occurs at about $r_{s}$$\simeq$25.
There is an experimental evidence \cite{ferro} for the ferromagnetic phase in 
Ca$_{1-x}$La$_{x}$B$_{6}$ at the density $r_{s}$=28 a.u., where the saturation 
moment of 0.07~$\mu_{B}$ per electron resists below the temperature 600~K, which is of the  order 
of the Fermi temperature of the electron gas.
However, for this experiment the iron substrate was chosen. Another experimental group 
\cite{Takagi} could not find any evidence for the intrinsic ferromagnetism in any of 
A$_{1-x}$La$_{x}$B$_{6}$ (A=Ca, Sr) samples.
As for the novel superconducting materials with triplet pairing, they cannot be described within a model 
based on the local spin density approximation, where the parametrization on a homogeneous gas results
works well. In order to describe those superconductors, like Sr$_{2}$RuO$_{4}$ or (TMTSF)$_{2}$X, it has been shown
by Shimahara in Ref.~\cite{Shimahara} that, one should add strong short range correlations to weaken the Coulomb
interaction and then the electron-phonon mechanism could cause pairing. Moreover a model assumed in our work
is 3D, while Sr$_{2}$RuO$_{4}$ should be described in 2D and (TMTSF)$_{2}$X in 1D.

There is also a question about an effect of the fluctuations which could mediate the pairing interaction 
(for such a model see the work of Abanov and Chubukov in Ref.~\cite{spin-fluct}).
Such calculations for the condensation energy in strongly correlated systems, where both effects of the 
electron-phonon and the electron-paramagnon interactions are taken into account, have been performed 
by Haslinger and Chubukov in Ref.~\cite{Chubukov-long}. 
The essential difference between those calculations and ours, if we included
phonons and paramagnons, would be in the frequency dependence of the gap function, which in our case is static
due to different formulation of the problem from the very begin.
For the inclusion of the dynamical effect to the gap, one needs to go beyond the Born-Oppenheimer approximation. 
In fact, such formulation exists within the multicomponent DFT scheme proposed by Kreibich and Gross \cite{MCDFT}, 
and developed by Van Leeuwen \cite{Robert}. 

In conclusion: we have calculated the condensation energy of the homogeneous electron gas at zero
temperature within  the density functional theory for superconductors. The random phase approximation
was assumed for the Coulomb interaction and no phononic contributions have been added. 
We did not consider pairing mechanism mediated by fluctuations.
Within this approach, there is no superconductivity for any momentum of the pairing potential 
for the densities up to $r_s$$\simeq$9. We found very weak superconductivity  for 
f-waves and higher-$l$ pairing at $r_s$=10, but this effect is so small that could be due to 
the neglecting of the vertex correction, or due to the accuracy of the numerical quadrature.

\begin{acknowledgments}
M.W. would like to thank 
Professor E.K.U.~Gross for suggesting the topic and hospitation
in W\"urzburg University and 
Hertie Foundation for a fellowship.
We are grateful to S. Yunoki and S. Kurth, M. L\"uders, M. Marques for helpful
discussions. The calculations were performed on the parallel computers
in the Lund's Centre for Scientific and Technical Computing, LUNARC, and 
in the Computing Centre of Karlsruhe University, and at CINECA in Bologna.
\end{acknowledgments}




\end{document}